\documentclass[12pt,superscriptaddress,nofootinbib,aps,floatfix,amsfonts,preprintnumbers,amsmath,amssymb,groupedaddress]{revtex4} 
\pdfoutput=1

\usepackage{graphicx, amsthm, multirow} 
\usepackage{url}
\usepackage{color}
\usepackage{dsfont}
\usepackage{stackrel}
\usepackage{bbm}
\usepackage{fancybox}
\usepackage{mdframed}
\usepackage{physics}
\usepackage{float}
\usepackage{bbold}
\usepackage{lpic}
\usepackage{array}
\usepackage{hyperref}

\def\37{$^{37}$Cl}
\def\71{$^{71}$Ga}
\def\8B{$^{8}$B}
\def\B7{$^{7}$Be}

\def\beq{\begin{equation}}
\def\eeq{\end{equation}}

\begin{document}

\title{\vspace*{1cm} 
Non-adiabatic Level Crossing in Resonant Neutrino Oscillations}

\author{Stephen J. Parke}
\affiliation{Theoretical Physics Department, \\ Fermi National Accelerator Laboratory,\\ Batavia, IL 60510, USA}

\begin{abstract} 
\vspace*{5mm}
Analytic results are presented for the probability of detecting an electron neutrino after passage through a resonant oscillation region. If the electron neutrino is produced far above the resonance density, this probability is simply given by {\mathversion{bold} $$\langle \,P_{\nu_e} \, \rangle  \approx \sin^2 \theta_0+ P_\text{\bf x} \cos 2 \theta_0\,,$$}where {\mathversion{bold} $\theta_0$} is the vacuum mixing angle. The probability is averaged over the production as well as the detection positions of the
neutrino and {\mathversion{bold}$P_\text{\bf x} $} is the Landau-Zener transition probability between adiabatic states. Finally, this result is applied to resonance oscillations within the solar interior.\\
\end{abstract}

\preprint{\href{https://inspirehep.net/literature/229055}{https://inspirehep.net/literature/229055} \\}
\preprint{
\href{https://doi.org/10.1103/PhysRevLett.57.1275}{DOI: 10.1103/PhysRevLett.57.1275}} 
\date{May 27, 1986 ---  FERMILAB-Pub-86/67-T}
\maketitle

Recently Mikheyev and Smirnov\cite{Mikheev:1986wj}  
and Bethe\cite{Bethe:1986ej} 
have revived interest in the solar-neutrino deficit by demonstrating that electron neutrinos produced in the sun can be efficiently rotated into muon neutrinos by passage through a resonant oscillation region. This mechanism may solve the solar-neutrino puzzle. In this paper, I present an analytic result for the probability of detecting an electron neutrino after passage through one or more resonant oscillation regions. This result is then used to show the regions of parameter space, the difference of the squared masses versus the vacuum mixing angle, for which the solar-neutrino puzzle is solved.

A neutrino state is assumed to be a linear combination of the two flavor states $\vert \nu_e \rangle$ and $\ket{\nu_\mu}$:
\beq
\ket{\nu(t)} = C_{e}(t)\ket{\nu_{e}} + C_{\mu}(t)\ket{\nu_\mu}.
\eeq
If the neutrinos are massive, then the mass eigenstates need not be identical to the flavor eigenstates, so that the Dirac equation which governs the evolution of the neutrino state is not necessarily diagonal in the flavor basis. This leads to the well known phenomena of vacuum neutrino oscillations. In the presence of matter, the non-diagonal nature of this evolution is further enhanced by coherent forward scattering which
can lead to resonant neutrino oscillations. Wolfenstein\cite{Wolfenstein:1977ue,Wolfenstein:1979ni}
has derived the Dirac equation for this process, in the ultra-relativistic limit, in terms of the vacuum mass eigenstates. Here, I use his result, in the flavor basis, after discarding a term proportional to the identity matrix, as this term only contributes an overall phase factor to the state $\ket{\nu(t)}$. The resulting Schrodinger-type wave equation is
\beq
i\frac {d}{dt}\left( \begin{array}{c} C_{e}\\ C_{\mu} \end{array}
\right)=
\frac{1}{2} \left( \begin{array}{cc}
-{\Delta_0}\cos2\theta_0 +\sqrt{2} G_F N &{\Delta_0} \sin2\theta_0 \\ 
{\Delta_0}\sin2\theta_0 &{\Delta_0} \cos2\theta_0-\sqrt{2} G_F N
\end{array} \right)
\left( \begin{array}{c} C_{e} \\ C_{\mu} \end{array}\right),
\eeq
where $\Delta_0= (m^2_2 - m^2_1)/2k$, ~~$m_1, \,m_2$ are
the neutrino masses, ~~k is the neutrino momentum, ~~$\theta_0$ is the vacuum mixing angle, $N$ is number density of electrons, and $G_F$ is the Fermi constant.
The constraints $\Delta_0>0$ and $\theta_0 <\pi/4$ are assumed. At an electron density, $N$, the matter mass eigenstates are
\begin{eqnarray}
        \ket{\nu_1,N} &~=~& \cos\theta_N~\ket{\nu_e} ~-~
\sin\theta_N~\ket{\nu_{\mu}} \,, \nonumber \\
        \ket{\nu_2,N} &~=~& \sin\theta_N~\ket{\nu_e} ~+~
\cos\theta_N~\ket{\nu_{\mu}}\,,
\end{eqnarray}
which have eigenvalues $\pm \Delta_N/2$, where
\beq
\Delta_N=[(\Delta_0\cos 2 \theta_0-\sqrt{2}G_F N)^2 +\Delta^2_0\sin^2 2 \theta_0]^{1/2} \,,
\eeq
and $\theta_N$ satisfies
\beq
\Delta_N \sin 2 \theta_N= \Delta_0 \sin 2 \theta_0 \,.
\eeq
These states evolve in time by the multiplication of a phase factor, if the electron density is a constant. For such a constant density there are three regions of interest: (i) Well below resonance, $\sqrt{2} G_F N \ll
\Delta_0 \cos 2 \theta_0$,
where the matter mixing angle is $\theta_N \sim \theta_0$ and the oscillation length is $L_0 = 2\pi/\Delta_0$. Typically, this is the region that the electron neutrinos are detected in. (ii) At resonance, 
$\sqrt{2}G_F N=\Delta_0 \cos 2 \theta_0$, where the matter mixing angle is $\theta_N=\pi/4$ and the resonant oscillation length is $L_R = L_0/\sin 2 \theta_0$, which for small vacuum mixing angle can be many times the vacuum oscillation length. (iii) Far above resonance, $\sqrt{2} G_F N \gg
\Delta_0 \cos 2 \theta_0$, where the matter mixing angle $\theta_N \sim \pi/2$, and the oscillation length 
$L_N = 2\pi /\Delta_N$ is much smaller than the vacuum oscillation length $L_0$. 
For the situation of current interest the electron neutrinos are produced above resonance, pass through resonance, and are detected in the vacuum.

If the electron density varies slowly, the states which evolve independently in time (the adiabatic states) are
$$\exp(-i \, \frac{1}{2} \int^{t} \Delta_N dt)~\ket{\nu_1,N(t)}$$
 and
$$\exp(+i \, \frac{1}{2} \int^{t} \Delta_N dt)~\ket{\nu_2,N(t)} \,. $$
 Therefore, it is convenient to
use these states, as the basis states, in the region for which there are
no transitions (away from the resonance region).
As a neutrino goes through resonance these adiabatic states
may be mixed, but on the
other side of resonance, the neutrino state can still be written as a
linear combination of these states.
That is, a basis state produced at time $t$, going through resonance at time
$t_r$, and detected at time $t^{\prime}$ is described by
\begin{eqnarray}
       &  \exp(-i \, \frac{1}{2} \int^{t}_{t_r} \Delta_N dt)~\ket{\nu_1,N(t)} \rightarrow
\nonumber \\ &
        ~a_1 ~\exp(-i \, \frac{1}{2} \int^{t^\prime}_{t_r}  \Delta_N dt)~\ket{\nu_1,N(t^{\prime})}
        ~+~ a_2 ~\exp(+i \, \frac{1}{2} \int^{t^\prime}_{t_r}  \Delta_N dt)~\ket{\nu_2,N(t^{\prime})} \,, \nonumber \\[4mm]
       & \exp(+i \, \frac{1}{2} \int^{t}_{t_r} \Delta_N dt)~\ket{\nu_2,N(t)} \rightarrow
\nonumber \\ &
         -a^{\ast}_2 ~\exp(-i \, \frac{1}{2}  \int^{t^\prime}_{t_r}  \Delta_N dt)~\ket{\nu_1,N(t^{\prime})}
        ~+~ a^{\ast}_1 ~\exp(+i \, \frac{1}{2}  \int^{t^\prime}_{t_r}  \Delta_N dt)~\ket{\nu_2,N(t^{\prime})} \,,
        \nonumber
\end{eqnarray}
\noindent
where $a_1$ and $a_2$ are complex numbers such that $|a_1|^2+|a_2|^2
=1$. The relationship between the coefficients, for these two
basis states, is due to the
special nature of the wave equation, Eq.(2).
The phase factors have been chosen
so that  coefficients $a_1$ and $a_2$ are characteristics of the transitions at
resonance and are not related to the production and detection of the
neutrino state.\\

Hence, the amplitude for producing, at time $t$, and
detecting, at time $t^{\prime}$, an electron neutrino after passage
through resonance is
\begin{eqnarray}
        & A_1(t)~\exp{-i  \, \frac1{2} \int^{t^{\prime}}_{t_r} \Delta_N dt}
~+~A_2(t)~\exp{+i \,  \frac1{2}\int^{t^{\prime}}_{t_r} \Delta_N dt} \,, & \nonumber
\end{eqnarray}
where
\begin{eqnarray}
A_1(t) &=&~\cos\theta_0~ \biggr[ a_1 \cos\theta_N ~\exp{+i \, \frac1{2} \int^{t}_{t_r} \Delta_N dt}
~-~ a^{\ast}_2 \sin\theta_N ~\exp{-i \, \frac1{2} \int^{t}_{t_r} \Delta_N dt}  \biggr]\,,
\nonumber \\[2mm]
A_2(t) &=& ~\sin\theta_0~\biggr[ a_2 \cos\theta_N ~\exp{+i  \, \frac1{2} \int^{t}_{t_r} \Delta_N dt}
~+~ a^{\ast}_1 \sin\theta_N ~\exp{-i \, \frac1{2} \int^{t}_{t_r} \Delta_N dt}  \biggr]. \nonumber
\end{eqnarray}
Thus the probability of detecting this neutrino as an electron
neutrino is given by
\begin{eqnarray}
         P_{\nu_e}(t,t^{\prime})&~=~& |A_1(t)|^2 ~+~ |A_2(t)|^2 ~+~
2|A_1(t)A_2(t)|\cos(\int^{t^{\prime}}_{t_r} \Delta_N dt ~+~ \Omega
) \nonumber
\end{eqnarray}
with $ \Omega = \arg(A_1^{\ast}A_2)$.
After averaging over the detection position, the detection averaged probability is given by
\begin{eqnarray}
        P_{\nu_e}(t) &=& {1 \over 2} + {1 \over 2}(|a_1|^2 ~-~ |a_2|^2)
\cos2\theta_N\cos2\theta_0 
-|a_1a_2| \sin2\theta_N \cos2\theta_0
 \cos(\int^{t}_{t_r} \Delta_N dt ~+~ \omega) \nonumber
\end{eqnarray}
with $\omega = \arg(a_1a_2)$. The last term shows that the phase
of the neutrino oscillation at the point the neutrino enters resonance
can substantially effect this probability.
Therefore, we must also average over the production position,
to obtain the fully averaged probability of detecting an electron neutrino as
\beq
 \mathversion{bold} \langle P_{\nu_{e}} \, \rangle  =~{1 \over 2} ~+~ \biggr({1 \over 2}~-~ P_x \biggr)
        \cos2\theta_N\cos2\theta_0 \,.
\eeq
where $P_x~=~ \vert a_2 \vert^2$, the probability of transition from $\vert \nu_2,N \rangle$
to $\vert \nu_1,N \rangle$ (or vice versa) during resonance crossing.
The adiabatic case \cite{Barger:1986ww}
is trivially obtained by setting $P_x=0$. 
Also, if the
electron neutrinos are produced at a density much greater than the
resonance density, so that $\cos2\theta_N \approx -1$, then
\beq
 \mathversion{bold} \langle P_{\nu_{e}} \, \rangle \approx ~ \sin^2\theta_0
        ~+~ P_x\,\cos2\theta_0 \,.
\eeq
Thus, in the very small $\theta_0$ limit, the survival
probability is just equal to the
probability of level crossing during resonance passage.\\

Similar calculations can also be performed for the case of double resonance
crossing (neutrinos from the far side of the Sun). Here we
must average not only over the production
and detection positions of the neutrino but also over the separation between
resonances. This sensitivity to the separation of the resonances
can be understood as
the effect of the phase of the oscillation as the neutrino enters the second
resonance region. The fully averaged probability of detecting an
electron neutrino is the same as Eq.(6) with $P_x$ replaced by
$P_{1x}(1-P_{2x}) ~+~ (1-P_{1x})P_{2x}$ (the classical probability
result).
Therefore, the generalization to any number of resonance regions, suitable
averaged, is obvious.\\
 
To calculate the probability, $P_x$, I make the approximation that the
density of electrons varies linearly in the transition region. That
is,
a Taylor series expansion is made about the resonance position and
the second and higher derivative terms are discarded;
\beq
        N(t)~\approx~ N(t_r)~+~(t-t_r){dN \over dt}|_{t_r}.
\eeq
In this approximation the probability of
transition between adiabatic states was calculated by Landau\cite{Landau:1932vnv} and
Zener\cite{Zener:1932ws}. 
This is achieved by solving the Schrodinger equation,
Eq.(2), exactly in this limit. The solution is in terms of Weber (parabolic cylinder) functions. Application of the Landau-Zenner result to the current situation gives
\beq
\mathversion{bold} P_{x} = \exp\left[-\frac {\pi}{2} 
\frac{\sin^{2}2\theta_0}{\cos2\theta_0}
\, \frac {\Delta_0}{| (1/N) dN/dt |_{t_r} }\right] \,.
\eeq
This expression, together with Eq.(6) are the main analytical results of this paper and  demonstrate that only the
electron number
density, at production, and the logarithmic derivative of this density, at
resonance,
determine the probability of detecting an electron neutrino in the vacuum. 
It should be emphasized here, that this result assumes that the
neutrino state is produced before significant transitions take place
and thus Eq.(9) is not valid for neutrinos produced in the transition
region.\\

From Eq.(9) the size of the transition region can be determined.
There are significant transitions ($P_x ~>~0.01$) if
$\theta_0 ~<~ \theta_{crit}$
where $\theta_{crit}$ satisfies
\beq
        {\sin^22\theta_{crit}\over \cos2\theta_{crit}}~=~
3~ {1 \over \Delta_0} ~\biggr|\frac{1}{N} \frac{dN}{dt} \biggr|_{t_r}.
\eeq
Hence, the maximum separation between
the eigenstates for which transitions take place is $\Delta_0
\sin2\theta_{crit}$. Therefore, the transition region is
defined by
\beq
        \Delta_N ~<~ \Delta_0 \sin2\theta_{crit}  .
\eeq
This can only happen if $\theta_0 < \theta_{crit}$.
In this transition region, the maximum variation of
the electron number density from the resonant value is $\pm \, \delta N $,
where
\begin{eqnarray}
        \delta N / N(t_r) = \sin2\theta_{crit}. \nonumber
\end{eqnarray}
Thus, the size of the transition region is
\begin{eqnarray}
        |t-t_r| &= & \sin 2\theta_{crit}
                \biggr/ \biggr| \frac{1}{N} \frac{dN}{dt} \biggr|_{t_r} . \nonumber
\end{eqnarray}
This is the maximum $|t-t_r|$ for which
the linear approximation must be good, so that Eq.(9) gives a
reasonable estimate of the probability of crossing. For an exponential
density profile, the Taylor series expansion is an expansion in
$\sin2\theta_{crit}$, so that small $\theta_{crit}$ is an
excellent approximation.\\
 
For the sun, the density profile is exponential except for the region near the center. In Fig. 1, I have plotted the probability contours for detection of an electron neutrino at the Earth in the $\Delta_0/\sqrt{2} G_F N_c$ vs $\sin 2\theta_0$
plane for an exponential density profile. $N_c$ is the electron number density at the point at which the electron neutrinos are produced. This plot depends only on the properties of the sun and this dependency is only through the combination $R_sN_c$ where $R_s$ is the scale height. For Fig. 1, I have used an $N_c$, corresponding to a density of 140 g/cm$^3$ and $Y_e$ =0.7. The scale height $R_s$ is 0.092 times the radius of the sun.\\

\begin{figure}[h]
  \centering
  \includegraphics[width=0.6\textwidth,angle=1.0]{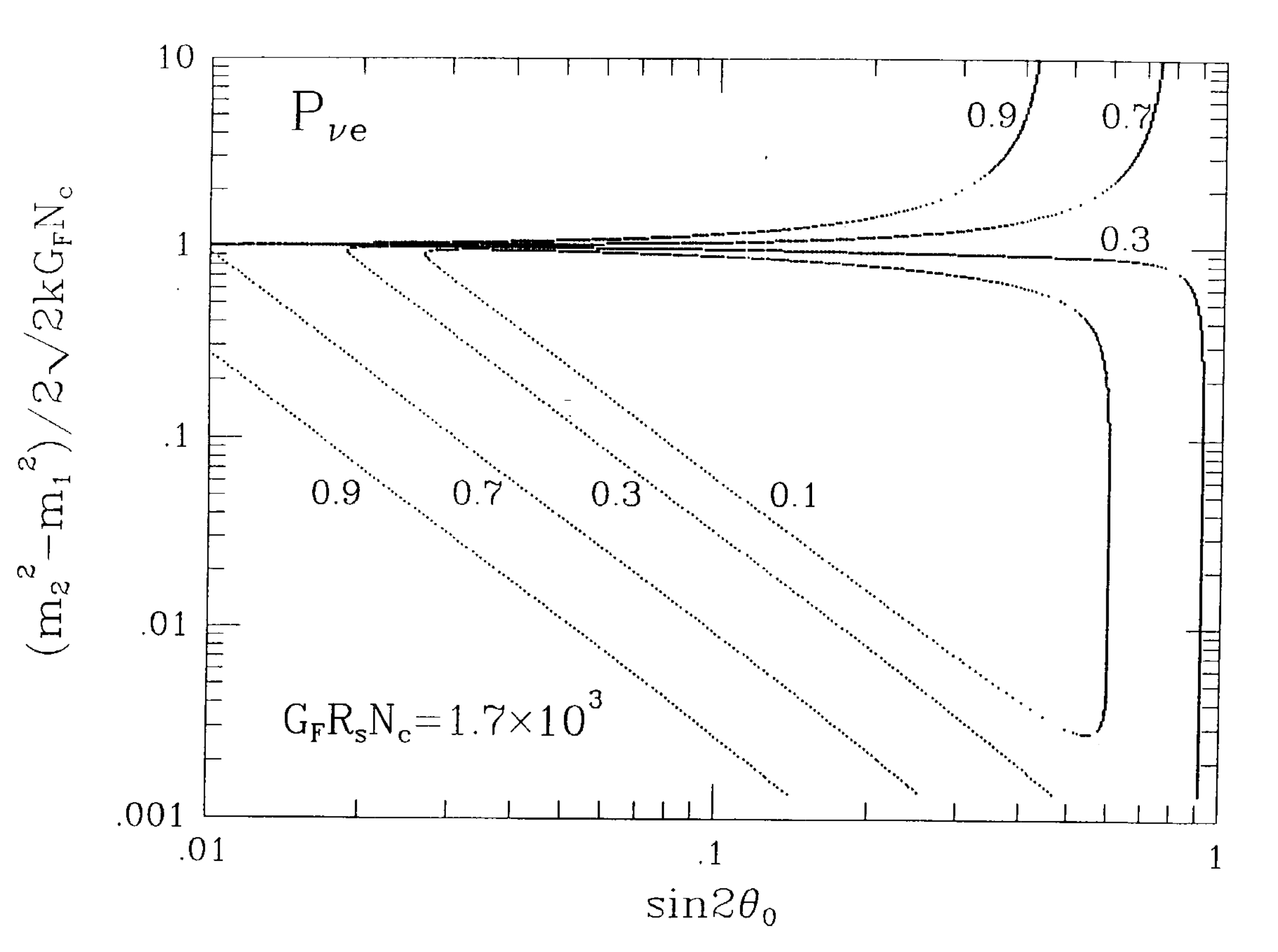}
    \caption{Probability-contour plot for detecting an electron neutrino at the Earth which was produced in the solar interior.
}
\end{figure}

Above the line  $\Delta_0 /\sqrt{2}G_F N_C=1/\cos 2 \theta_0$, the neutrinos never cross the resonance density on their way out of the sun. Here, the probability of detecting an electron neutrino is close to the standard neutrino oscillation result. Below this line, the effects of passing through resonance come into play. Inside the 0.1 contour, there is only a small probability of transitions between the adiabatic states as the neutrino passes through resonance. To the right-hand side of this contour, the probability of detecting a neutrino grows, not because of transitions, but because both adiabatic states have a substantial mixture of electron neutrino at zero density. To the left-hand side and below this contour, the probability grows because here there are significant transitions between the adiabatic states as the neutrino crosses resonance. The diagonal lines of these contours have slope of -2 because of the form of $P_x$. It is only the intercept of these lines which depends on the product $R_S N_c$. Therefore, if one wishes to change the production density, which is held fixed in this plot, only these lines need to be shifted. 
In fact, a line labeled with $P_e$ ``crosses''  $\Delta_0 /\sqrt{2}G_F N_C=1$, when a small $\theta_0$ satisfies
\beq
\frac{\sin^2 2 \theta_0}{\cos 2 \theta_0} = \frac{-\sqrt{2}\, \text{ln}(P_e)}{\pi G_F R_s N_c} \,.
\eeq
Note that I find the probability of detecting an electron
neutrino, which crosses resonance, to be greater than 0.25 when $\theta_0 < 0.01$.\\

This iso-probability plot can easily be converted into an approximate iso-SNU (solar neutrino units) plot for the Davis {\it et al}  experiment\cite{Bahcall:1985bda}. 
The predicted result for this experiment\cite{Bahcall:1981zh} 
is 6 SNU, with 4.3 SNU coming from the \8B neutrinos and 1.6 SNU from the lower-energy neutrinos (pep, \B7, $^{13}$N, and $^{15}$O), whereas Davis {\it et al}  observe 2.1$\pm$0.3 SNU. Roughly speaking, the 2 SNU contour, in the $m^2_2-m^2_1$ vs $\sin2 \theta_0$ log-log plot, will be a triangle, similar to the 0.3 contour of Fig. 1, with rounded corners. The three straight sections of this triangle are approximately given below. The horizontal line is given by choosing the parameters so that all the low-energy neutrinos and only 12\% of the \8B neutrinos are observed. This gives the constraints obtained by Bethe\cite{Bethe:1986ej},  
\begin{eqnarray}
& (m^2_2 - m^2_1 ) \approx 8 \times 10^{-5} eV^2 \,, \notag  \\
& 0.03 < \sin 2\theta_0 < 0.6\, .
\end{eqnarray}
For the vertical line, the probability of detecting an electron neutrino)is nearly independent of energy, if
$1> \Delta_0/\sqrt{2}G_FN > 10^{-3}$. Therefore, we need to
reduce all neutrinos by 30\%\,\cite{Barger:1986ww}. 
This is achieved when
\begin{eqnarray}
& 8 \times 10^{-8} eV^2 < (m^2_2 - m^2_1 ) < 1 \times 10^{-5} eV^2 \,, \notag  \\
& \sin 2\theta_0 \approx 0.9 .
\end{eqnarray}
For the diagonal line, we need to arrange that the Davis experiment only observed 50\% of the \8B neutrinos and none of the lower-energy neutrinos\cite{Rosen:1986jy, Kolb:1986nd}.  
This is achieved when the probability for the mean \8B neutrino, weighted by the detector cross section (energy
$\sim$10~MeV), is 0.5. This gives the following constraint:
\begin{eqnarray}
&(m^2_2 - m^2_1 )\sin^2 2\theta_0 =3 \times 10^{-8} eV^2 \,, \notag  \\
&0.03 < \sin 2\theta_0 < 0.6.
\end{eqnarray}
To summarize, Eqs. (13)—(15) give regions of parameter space for which the expected result from the Davis experiment is $\sim$2 SNU.\\

Since the proposed gallium experiment observes lower-energy neutrinos, from the pp process, these three regions will be distinguishable by use of the results of this experiment. More precise iso-SNU plots, for both experiments, are being generated taking into account the production energy and production position distributions of the neutrinos from the various processes within the solar interior. \\

I would like to acknowledge discussions with T. Walker and R. Kolb. Fermilab is operated by the Universities Research Association Inc., under contract with the United States Department of Energy.\\

 \newpage

\bibliography{PRL86a}

\end{document}